# Recommendation of Scholarly Venues Based on Dynamic User Interests

**Hamed Alhoori [a,b,1] and Richard Furuta [c]**

[a] Northern Illinois University, DeKalb, IL, USA
[b] Argonne National Laboratory, IL, USA
[c] Texas A&M University, College Station, TX, USA

## Abstract

The ever-growing number of venues publishing academic work makes it difficult for researchers to identify venues that publish data and research most in line with their scholarly interests. A solution is needed, therefore, whereby researchers can identify information dissemination pathways in order to both access and contribute to an existing body of knowledge. In this study, we present a system to recommend scholarly venues rated in terms of relevance to a given researcher's current scholarly pursuits and interests. We collected our data from an academic social network and modeled researchers' scholarly reading behavior in order to propose a new and adaptive implicit rating technique for venues. We present a way to recommend relevant, specialized scholarly venues using these implicit ratings that can provide quick results, even for new researchers without a publication history and for emerging scholarly venues that do not yet have an impact factor. We performed a large-scale experiment with real data to evaluate the current scholarly recommendation system and showed that our proposed system achieves better results than the baseline. The results provide important up-to-the-minute signals that compared with post-publication usage-based metrics represent a closer reflection of a researcher's interests.

---

[1] Corresponding author. E-mail address: alhoori@niu.edu (H. Alhoori)

## 1. Introduction

In addition to the variety of challenges researchers face from the rising number of scholarly events and venues, the important task of identifying relevant publication opportunities is further complicated due to the expansion and overlap of what were previously discrete specializations. More and more collaboration is taking place between disciplines in the research landscape, which is leading to decreased compartmentalization overall. Increasingly complex academic sub-disciplines and emerging interdisciplinary research areas, though certainly a net gain for the community as a whole, compound this problem. In such a sophisticated research environment, researchers are finding it challenging to remain up to date on new findings, even within their own disciplines (Kuruppu & Gruber, 2006; Murphy, 2003). Furthermore, "context-drift" in scholarly communities is becoming more prevalent as researchers expand, evolve, or adapt their interests in rapidly changing subject areas.

Generally, researchers become aware of scholarly venues related to their research interests by word of mouth from lab members, departmental colleagues, and members of other scholarly communities; through online searches for scholarly material; and from rankings of venues and publishers' reputations (Buchanan, Cunningham, Blandford, Rimmer, & Warwick, 2005; Chu & Law, 2007). In the past, these approaches have yielded satisfactory results, as there were relatively few venues related to any given field. However, in today's multifaceted, diverse, and interdisciplinary scholarly environment, researchers can become acquainted with newly available and relevant specialized venues only by spending considerable time and effort explicitly searching for venues that align with their research interests.

It is also essential for funding agencies to become aware of new avenues of research across fields in order to determine future allocations. Further, new interdisciplinary research areas lead to greater challenges for research institutes as they strive to understand dynamic information needs and information-seeking behaviors. Information specialists need prompt and seamless measurements of researchers' readings in order to make decisions on venue subscriptions, instead of relying blindly on the venue's impact factor or on users' explicit requests. For example, Springer provides its users with a form for recommending journals to librarians (Springer, 2015), but this feedback represents only the interests of the individuals who submit recommendations, rather than providing a picture of the entire constituency's needs.

Many rankings of scholarly venues have been created and used to help researchers become more aware of specific scholarly communities. However, knowing that very prestigious journals, such as *Science* and *Nature,* are considered top venues for multidisciplinary fields does not help researchers seeking more specialized venues and communities. Moreover, traditional citation analysis cannot provide quick, adaptive results, especially for new scholarly venues that do not yet have an impact factor.

A number of online services provide collections of venues in an attempt to alleviate some of these problems. For example, the HCI Bibliography (Perlman, 1991) is a specialized bibliographic database on Human-Computer Interaction. AllConferences[2] and Lanyrd[3] are global conference and event directories. ConferenceAlerts,[4] EventSeer,[5] and WikiCFP[6] provide notifications of upcoming academic events based on keywords. ConfSearch (Kuhn & Wattenhofer, 2008) enables researchers to search for computer science conferences using keywords, related conferences, and authors. ConfAssist (Singh, Chakraborty, Mukherjee, & Goyal, 2016) classifies conferences as top-tier or not.

---

[2] http://www.allconferences.com/
[3] http://lanyrd.com/

However, in this era of big data, retrieving relevant results by manually searching and browsing online is no longer the only approach to discover new information, not is it generally the most efficient approach. Studies have been conducted in an effort to offer techniques capable of accelerating scholarly discovery, such as summarization, visualization (Gove, Dunne, Shneiderman, Klavans, & Dorr, 2011), and collaborative information synthesis (Blake & Pratt, 2006). Recommender systems have been introduced to filter the overwhelming amount of data by using various data analysis techniques to alleviate information overload (Shenk, 1997; Speier, Valacich, & Vessey, 1999). Recommender systems are already entrenched in the digital landscape, as they provide millions of online users with continually updated suggestions for news, books, restaurants, tourism, movies, and television programs.

With the proliferation of publications, researchers are utilizing academic social networks and reference management systems in order to find, store, and manage references (Farooq, Song, Carroll, & Giles, 2007). Social and online reference management systems enable users to bookmark references to research content, as well as tag, review, and rate research content within their profiles. Scholarly tools such as these play an essential role in the organization of personal article collections and the generation of bibliographies across the research landscape today. Scholarly communities are sharing these digital reference libraries, and this open sharing encourages the formation of new research groups. Such online personal collections or repositories also accurately reflect researchers' current and past reading, and indicate changes in their interests over time, making these datasets prime targets for recommendation analytics.

---

[4] http://www.conferencealerts.com
[5] http://eventseer.net/
[6] http://www.wikicfp.com/

In previous work (Alhoori, 2016; Alhoori & Furuta, 2011), we found that several of the participating researchers expressed a notable desire to be aware of new and well-established scholarly venues and events related to their shifting research interests. In this paper, we build a personal measure for evaluating venues based on user-centric altmetrics and analysis of readings, rather than relying on conventional citation-based metrics. Then, we augment the researchers' awareness and recommend semantically related scholarly venues based on their interests. In creating this measure, we draw on data from CiteULike,[7] a well-known social reference management system.

This paper is structured as follows: In section 2, we discuss related work. In section 3, we describe an approach for measuring an implicit rating for scholarly venues by monitoring researchers' behavior. In section 4, we explain the data collection and the experiments. In section 5, we present and discuss the results.

## 2. Related Work

Recommender systems streamline and augment a person's decision-making process, especially when inadequate information is available with which to make an informed decision (Resnick & Varian, 1997). One well-known recommender technique is collaborative filtering (CF) (Resnick, Iacovou, Suchak, Bergstrom, & Riedl, 1994; Schafer, Frankowski, Herlocker, & Sen, 2007). User-based collaborative filtering stems from the idea that users whose respective ratings show a high level of agreement and/or who have a similar history of behaviors are likely to continue to show agreement in these regards. This algorithm searches for users who share similar patterns to those of a current user and uses their ratings to predict unidentified preferences for the current user. Item-based collaborative filtering uses similarities between item ratings to predict users' preferences instead of using similarities between users' ratings (Sarwar, Karypis, Konstan, & Reidl, 2001).

---

[7] http://www.citeulike.org/

Other recommender systems use a matrix factorization approach[8] based on the stochastic gradient descent (SGD) (Bottou & Bousquet, 2008), singular value decomposition (SVD) (Badrul Sarwar, Karypis, Konstan, & Riedl, 2000), or SVD++ (Koren, Bell, & Volinsky, 2009). SGD is an iterative learning algorithm for minimizing the error between actual and predicted ratings. SVD reduces the dataset by eliminating insignificant users or items. SVD++ constitutes an improvement on SGD in which it not only considers ratings but also considers who has rated what (e.g., rating an item is an indication of preference).

Recommender systems have been used to recommend movies (Wu & Niu, 2015), research papers (Beel, Gipp, Langer, & Breitinger, 2015), collaborators (Yan & Guns, 2014), experts (Protasiewicz et al., 2016), reviewers (Basu, Cohen, Hirsh, & Nevill-Manning, 2001), citations (Caragea, Silvescu, Mitra, & Giles, 2013), and tags (Song, Zhang, & Giles, 2011).

---

[8] https://mahout.apache.org/users/recommender/matrix-factorization.html

The need to connect authors and readers goes back to at least 1974 when Kochen and Tagliacozzo (1974) proposed a service to suggest journals for authors' manuscripts using a mathematical model that took into consideration relevance, acceptance rate, circulation, prestige and publication lag. However, until just a few years ago very little progress had been made in this area. Since then, due in large part to the increasing information overload researchers face when searching for new venues, there has been a resurgence in research and development surrounding the recommendation of scholarly events (Huynh & Hoang, 2012).

Klamma et al. (2009) recommended academic events based on researchers' event participation history, whereas (H. Luong, Huynh, Gauch, Do, & Hoang, 2012; H. P. Luong, Huynh, Gauch, & Hoang, 2012) used co-authors' publication history to recommend venues. Boukhris and Ayachi (2014) proposed a hybrid recommender for upcoming conferences related to computer science based on venues from co-authors, co-citers, and co-affiliated researchers. Pham et al. (2011) clustered users on social networks and used the number of papers a researcher had published in a venue to derive the researcher's rating for that venue. eTBLAST (Errami, Wren, Hicks, & Garner, 2007) and the Journal Article Name Estimator (Jane) (Schuemie & Kors, 2008) recommend biomedical journals based on an assessment of abstract similarity. Silva et al. (2015) considered the quality and relevance of manuscripts in order to recommend journals. They also analyzed the authors' social networks and identified journals in which similar researchers had published. Other venue recommendation approaches have based ratings on the topic and writing style of a paper (Yang & Davison, 2012), the title and abstract of a paper (Medvet, Bartoli, & Piccinin, 2014), an analysis of PubMed log data (Lu, Xie, & Wilbur, 2009), and personal bibliographies and citations (Küçüktunç, Saule, Kaya, & Çatalyürek, 2012; Küçüktünc, Saule, Kaya, & Çatalyürek, 2013).

Recently, some online services have started to provide support for locating relevant journals using title, keyword, and abstract matching. These services include Elsevier Journal Finder (Kang, Doornenbal, & Schijvenaars, 2015),[9] Springer Journal Selector,[10] EndNote Manuscript Matcher,[11] Jane,[12] and Edanz Journal Selector[13].

In addition, more research has been carried out in recent years on recommending events in general. For example, Xia et al. (2013) presented a socially aware recommendation system for conference sessions, and Quercia et al. (2010) used mobile phone location data to recommend social events. Minkov et al. (2010) proposed an approach to recommending future events, whereas Khrouf and Troncy (2013) used hybrid event recommendations with linked data.

Most research on scholarly venue recommendation to date has used citation analysis and the publication or participation history of researchers to build recommendations. Unfortunately, this model cannot be widely generalized, as it would not be useful for new researchers or graduate students who lack an established record of scholarly activity. Furthermore, using only the venues in which a researcher has previously published work undermines the recommendation process, as a researcher might be interested in new research areas in which she or he has not yet published any articles. This research study explores pathways with the purpose of drawing on a researcher's current personal article collections and readings to build tailored venue recommendations.

## 3. Personal Venue Rating (PVR)

---

[9] http://journalfinder.elsevier.com/
[10] http://www.springer.com/gp/authors-editors/journal-author/journal-author-helpdesk/preparation/1276
[11] http://endnote.com/product-details/manuscript-matcher
[12] http://jane.biosemantics.org/
[13] https://www.edanzediting.com/journal-selector

Venues can prove difficult to analyze for the purpose of building recommendations, as explicit metadata or user ratings are scarce and hard to come by. Research articles, however, have a variety of associated metadata fields. References in a researcher's library can thus provide easily accessible, indirect information pertaining to a researcher's interests. We used references and the years in which they were added to a researcher's library in the measurement of personal venue rating. *PVR* takes into consideration how a researcher's interest in a given venue has changed over time. In Equation 1, we define *PVR* as a weighted sum for researcher $u$ and venue $v$, and we refer to it as $r_{u,v}$ :

$$r_{u,v} = \sum_{i=y}^{1} w \log(v_{u,i} + 1) \tag{1}$$

$v_{u,i}$ denotes the number of references in a researcher's library $u$, from a specific venue $v$, which the researcher added during a certain year of the total number of years $y$ during which the researcher followed venue $v$. The weight $w$ increases the importance of newly added references and is equal to $i$. *PVR* favors researchers who have followed a venue for several years over researchers who have added numerous references from a venue over fewer years. The *log* minimizes the effect of adding numerous references and helps to reduce the impact of potential shilling attempts (Lam & Riedl, 2004). The addition of one allows for the case of one reference to be added to a library in a year. We used the year that a reference was added to the researcher's library, as it is more relevant to the researcher than the year the article was published.

## 4. Data and Experiments

### 4.1 Metrics

We conducted an offline experiment using the CiteULike dataset,[14] which consists of a CiteULike article ID, username, date, and time an article was added, as well as tags applied to the article. We used the article ID to crawl the CiteULike website, and we randomly downloaded 554,023 metadata files. These files contained more details about each article, including a link to the publisher's website, a list of authors, an abstract, a DOI, BibTex bibliographic information, the venue name, the year of publication, and a list of the users who had added that article to their personal article collections. We used an XML parser to clean and extract information from the CiteULike files. Using the BibTex field, we selected only the files that included either conference or journal data. Our final dataset contained 407,038 files. We then extracted the venue details from each article and collected a total of 1,317,336 postings of researcher–article pairs, as well as 614,361 researcher–venue pairs (Alhoori & Furuta, 2013). Figure 1 shows the PVR recommendation system architecture.

---

[14] http://www.citeulike.org/faq/data.adp

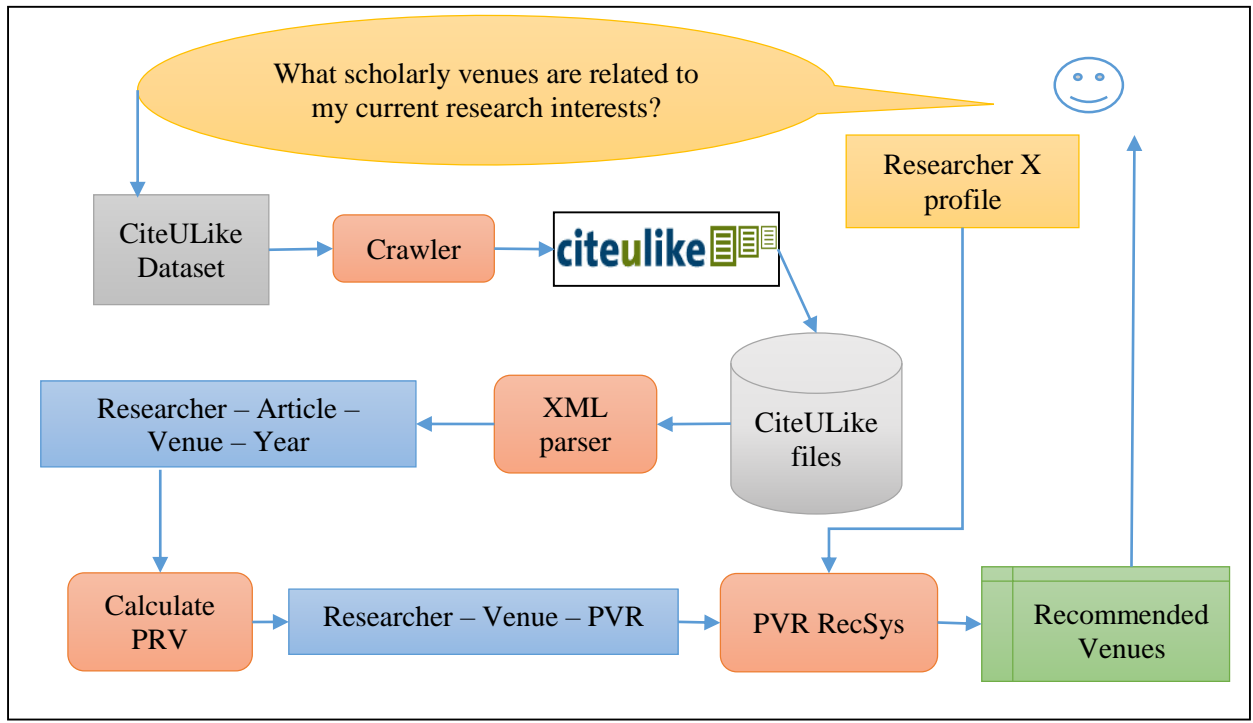

**Figure 1.** The PVR recommendation system architecture

We implemented user-based collaborative filtering (CF), item-based CF, stochastic gradient descent (SGD), and singular value decomposition (SVD++) algorithms from Apache Mahout (Apache Software Foundation, 2015) to recommend venues to researchers. We compared researchers with similar interests in terms of their PVRs. In recommendation systems, we need to find similarities between users or items. We choose to apply cosine similarity, Pearson correlation similarity, and Euclidean distance similarity in this study, because they are widely used in this type of analysis (Herlocker, Konstan, Borchers, & Riedl, 1999). The cosine similarity between two vectors is the angle between them and is usually useful for sparse data. The smaller the angle is to zero then the more similar those two vectors are. The Pearson correlation shows that when a series of ratings increase or decrease together. It is considered a centered version of cosine similarity (e.g., a cosine similarity when the two vectors have a mean of zero). The Euclidean distance similarity utilizes the distance between two vectors to calculate the distance between users.

The cosine similarity ($sim_{x,u}$) between a researcher $x$ and another researcher $u$ was computed as shown in Equation 2, where $\vec{x}$ and $\vec{u}$ are two vectors representing the ratings of the two researchers, $||\vec{u}||$ is the vector's Euclidian length, and $n$ is the number of venues rated by both researchers. The cosine similarity is the cosine angle between them.

$$sim_{x,u} = \cos\ (\theta) = \frac{\vec{x} \cdot \vec{u}}{||\vec{x}|| \times ||\vec{u}||} = \frac{\sum_{v=1}^{n}(r_{x,v})(r_{u,v})}{\sqrt{\sum_{v=1}^{n}(r_{x,v})^2}\ \sqrt{\sum_{v=1}^{n}(r_{u,v})^2}} \tag{2}$$

The Pearson correlation similarity ($sim_{x,u}$) is measured by Equation 3. $\bar{r_u}$ is the average $PVR$ for researcher $u$.

$$sim_{x,u} = \frac{\sum_{v=1}^{n}(r_{x,v} - \bar{r_x})(r_{u,v} - \bar{r_u})}{\sqrt{\sum_{v=1}^{n}(r_{x,v} - \bar{r_x})^2}\ \sqrt{\sum_{v=1}^{n}(r_{u,v} - \bar{r_u})^2}} \tag{3}$$

Equation 4 shows the Euclidean distance. $V_{x,u}$ is the set of venues rated by both $x$ and $u$.

$$Euclidean\ distance(x,u) = \sqrt{\frac{\sum_{v\ \in\ V_{x,u}}(r_{x,v} - r_{u,v})^2}{|V_{x,u}|}} \tag{4}$$

In the Euclidian distance similarity algorithm, a greater distance indicates fewer similar researchers. We therefore used $(1/(1 + distance)$ to identify similar researchers. Active researchers would have many similar venues with other researchers, which would create high correlations based on just a few co-rated venues. Therefore, for researchers who shared fewer co-rated venues than a threshold, we applied significance weighting (Herlocker et al., 1999), which reduced the overestimated similarity weight by a factor proportional to the number of co-rated venues. However, for researchers who shared more venues than the threshold, we did not adjust the similarity weight, as the more venues that are shared the more reliable the similarity. We found that this technique improved the accuracy of our results.

Users tend to assign different ranges of ratings. In other words, some users may generally assign high ratings whereas others may generally assign low ratings. Therefore, we normalized the ratings using a user mean-centering prediction. Prediction $p_{x,v}$ for an active user $x$ and for venue $v$ is measured by Equation 5. $\overline{r_x}$ is the average rating assigned by user $x$ to all the rated items. $U_v(x)$ is the set of user $x's$ neighbors (similar users) who rated venue $v$. $\overline{r_u}$ is the average rating for user $u$ for the items rated by both $x$ and $u$ (i.e., all the co-rated items).

$$p_{x,v} = \overline{r_x} + \frac{\sum_{u \in U_v(x)} (r_{u,v} - \overline{r_u}) sim_{x,u}}{\sum_{u \in U_v(x)} |sim_{x,u}|} \qquad (5)$$

We also calculated the item mean-centering prediction, as shown in Equation 6. $\overline{r_v}$ is the average rating of venue $v$ for all users. $W_x(v)$ is the set of venues similar to venue $v$ and rated by user $x$ (venues rated by $x$ as most similar to $v$). $\overline{r_w}$ is the average rating for venue $w$ derived from the ratings of all the users who rated venues $w$ and $v$.

$$p_{x,v} = \overline{r_v} + \frac{\sum_{w \in W_x(v)} (r_{x,w} - \overline{r_w}) sim_{v,w}}{\sum_{w \in W_x(v)} |sim_{v,w}|} \qquad (6)$$

### 4.2 Evaluation Metrics

We used a Boolean recommendation as a baseline and compared it with recommendations for scholarly venues based on PVR implicit ratings. Boolean ratings assume that all venues added by researchers are good venues and receive the highest rating. In the case of Boolean ratings, we used the log-likelihood similarity algorithm (Dunning, 1993). To rank the Boolean recommendations, venues affiliated with a large number of similar users were weighted more heavily (Owen, Anil, Dunning, & Friedman, 2011).

To measure the recommendations' performance, we measured precision, recall, and normalized discounted cumulative gain (NDCG) (Järvelin & Kekäläinen, 2002; McNee, Riedl, & Konstan, 2006). Precision is derived by dividing the number of relevant venues recommended according to the researcher's interests by the number of recommended venues, as shown in Equation 7. Recall is derived by dividing the number of relevant venues recommended by the number of relevant venues, as shown in Equation 8. For each user, the top 10 venues ranked by PVR were removed and the percentage of those 10 venues that appeared in the proposed top recommendations constituted the precision at 10 (P@10).

$$Precision = \frac{|relevant\ venues \cap top\ venues|}{|top\ venues|} \tag{7}$$

$$Recall = \frac{|relevant\ venues \cap top\ venues|}{|relevant\ venues|} \tag{8}$$

Discounted cumulative gain (DCG) measures the extent to which a venue ranking is relevant to a user's ideal ranking, as shown in Equation 9.

$$DCG_p = \sum_{v=1}^{p} \frac{2^{rel_v} - 1}{log_2(1 + v)} \tag{9}$$

$rel_v$ is the relevance assigned by a researcher to the venue at position $p$. We measured the normalized discounted cumulative gain (NDCG), which ranges from 0.0 to 1.0, with 1.0 as the ideal ranking, as shown in Equation 10. As recommendation lists vary in length, we used NDCG. $IDCG_p$ is the maximum possible ideal DCG at position $p$.

$$NDCG_p = \frac{DCG_p}{IDCG_p} \tag{10}$$

We also incorporated user coverage (Good et al., 1999;  Herlocker, Konstan, Terveen, & Riedl, 2004; Sarwar et al., 1998), which is the percentage of users for whom the system was able to recommend venues. Additionally, we tested for the normalized mean absolute error (NMAE) and the normalized root mean square error (NRMSE), which are independent rating scales. MAE (Shani & Gunawardana, 2009), the absolute deviation of a researcher's predicted PVR and observed PVR, is calculated as shown in Equation 11. $p_{u,v}$ is the predicted rating for venue $v$, and $r_{u,v}$ is the actual rating.

$$MAE = \frac{\sum_{v=1}^{n} |p_{u,v} - r_{u,v}|}{n} \tag{11}$$

RMSE is measured using the square root of the average squared difference between a researcher's predicted PVR and observed PVR as shown in Equation 12.

$$RMSE = \sqrt{\frac{\sum_{v=1}^{n} (p_{u,v} - r_{u,v})^2}{n}} \tag{12}$$

We used 70% of the data as a training set and 30% as a test set. We selected recommendations by choosing a threshold per user that was equal to the user's average $PVR$.

## 5.  Results and Discussion

We began by comparing user similarities with and without significance weighting. Collaborative filtering is affected by the cold-start problem (Schein, Popescul, Ungar, & Pennock, 2002) in which the system cannot produce good recommendations for new researchers or unrated venues. As our dataset includes thousands of venues and as each researcher would only know few of them, many of the venues will not have a rating from any given researcher. In addition, the Pearson correlation was not able to compute any similarity between users in some cases. For example, researchers who had added only one reference to their libraries and no other researchers share the same or similar reference. Therefore, venues that did not have an indirect rating were inferred to have an average rating.

Figure 2 shows that the use of significance weighting improved the accuracy, recall, and NDCG. The use of inferred ratings showed some improvement in the results as the neighborhood size increased. Further, Figure 2 also shows that Euclidean-weighting achieved better precision, recall, NDCG, and users' coverage than Pearson-weighting. We tested other similarities such as Spearman correlation similarity and Tanimoto coefficient similarity, but we found that Pearson correlation similarity and Euclidean distance similarity achieved better results.

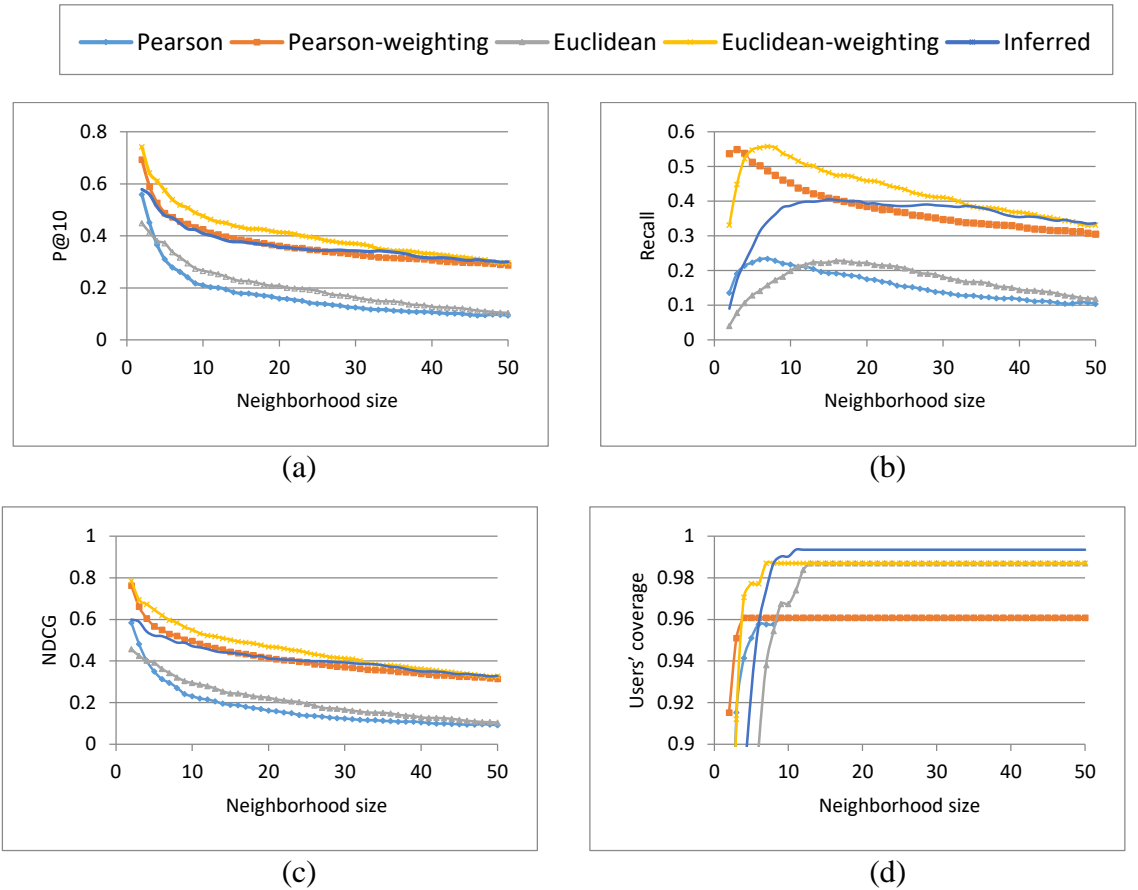

**Figure 2.** Comparison of the user-based CF algorithm with different similarities and neighborhood sizes

We then compared similarities that used PVR ratings and the user-based collaborative filtering algorithm with the baseline Boolean recommendation, as shown in Figure 3. Figure 3 (a–c) demonstrates that in general the PVR implicit ratings achieved higher precision, recall, and NCDG at lower neighborhood sizes. This result shows that in multidisciplinary environments such as CiteULike, researchers with a high level of similarity tend to cluster in small groups, which explains why the baseline achieved better results when the number of similar researchers increased (neighborhood size). Figure 3 (d) shows the users' coverage and that the PVR model was able to provide recommendations for up to 98% of users.

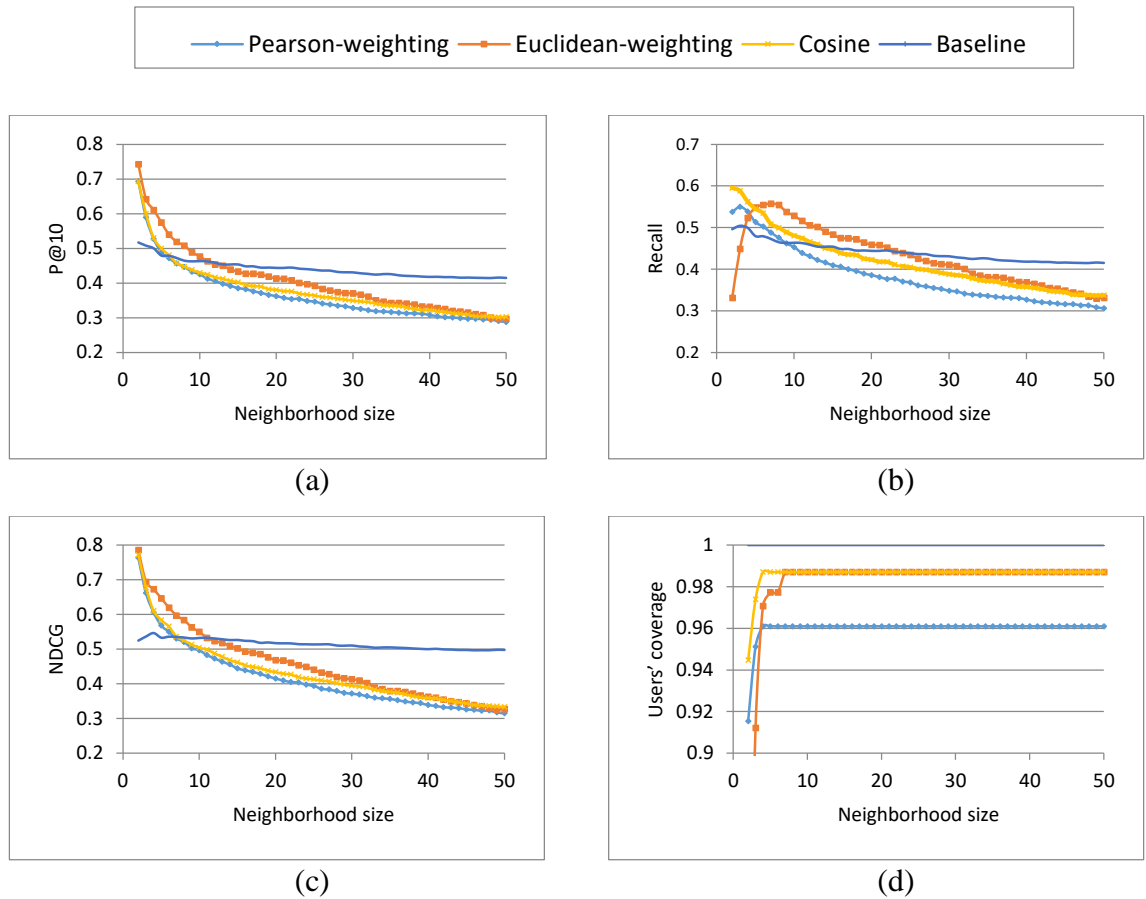

**Figure 3.** Comparison of the user-based CF algorithm with similarities that use PVR ratings and the baseline at different neighborhood sizes

Figure 4 illustrates the use of thresholds for researchers who are at least T percentage similar instead of fixed neighborhood sizes. Pearson-weighting achieved the highest P@10 and the highest NDCG, whereas the Boolean recommendations achieved the highest recall and the highest coverage at low thresholds. Figure 4 (a)(c) shows that the threshold increases and reaches its maximum at a value around 0.6. Then it starts to drop since the neighborhood does contain fewer researchers.

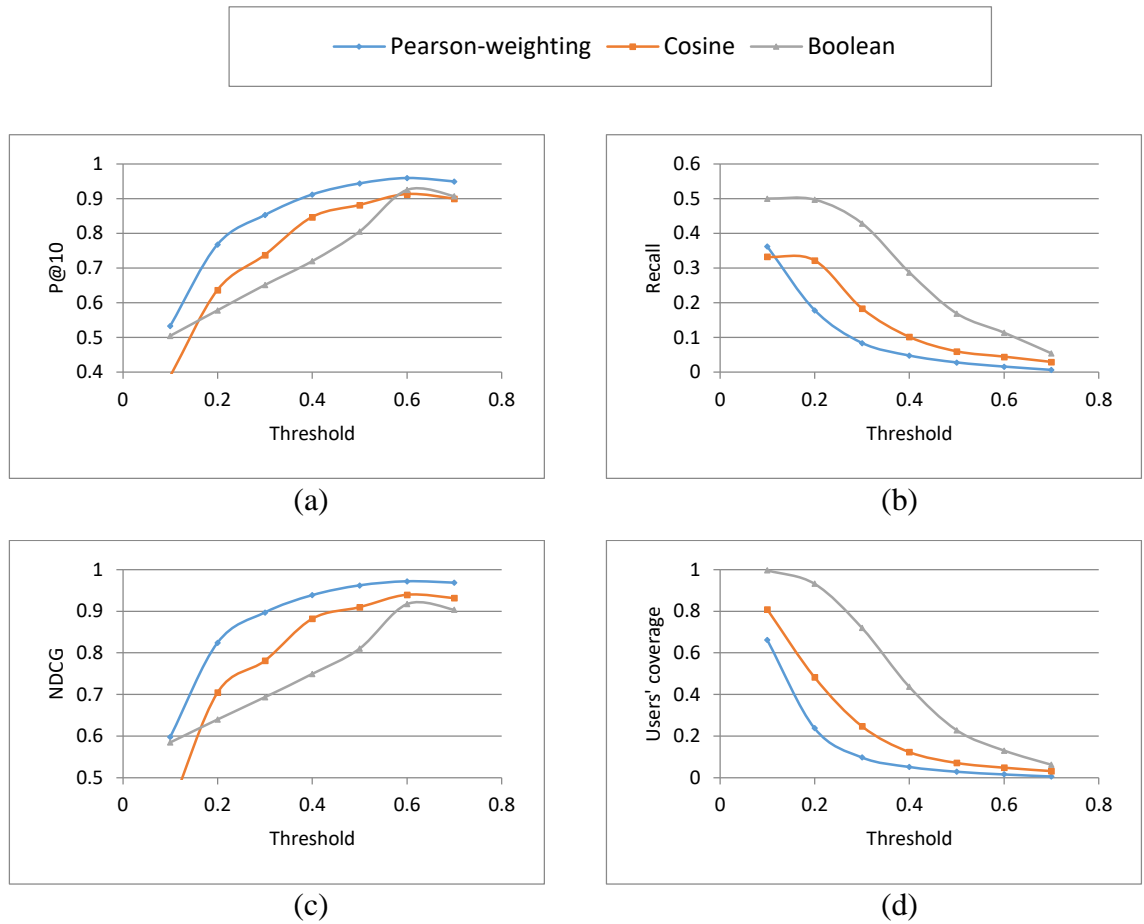

**Figure 4.** Comparison of user-based CF performance using different similarities and thresholds

We measured NMAE and NRMSE at different neighborhood sizes as Figure 05 shows, and found that the Euclidean-weighting achieved the lowest NMAE as well as the lowest NRMSE. Figure 5 also shows that Pearson-weighting and cosine similarity result in the highest error. Figures 2, 3, and 5 show that Euclidean-weighting achieved the highest level of accuracy and the lowest level of errors.

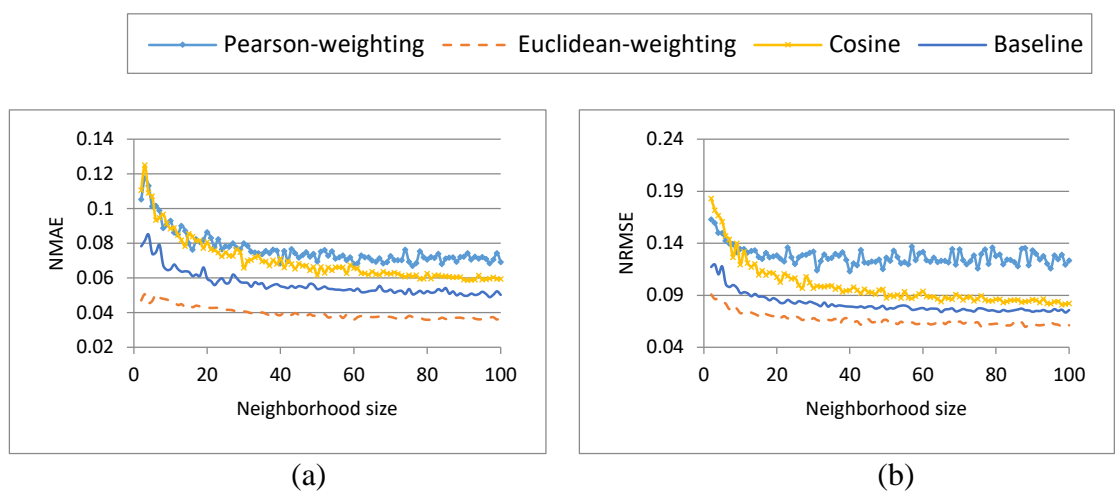

(a)                                    (b)

**Figure 5.** NMAE and NRMSE for user-based CF with different neighborhood sizes

We compared the performance of four algorithms that used PVR ratings at different percentages of the training set (Figure 6), and we found that SVD++ achieved the lowest NMAE and the lowest NRMSE. Item-based CF performed better than User-based CF, which shows that researchers' interests and implicit ratings are changing over time.

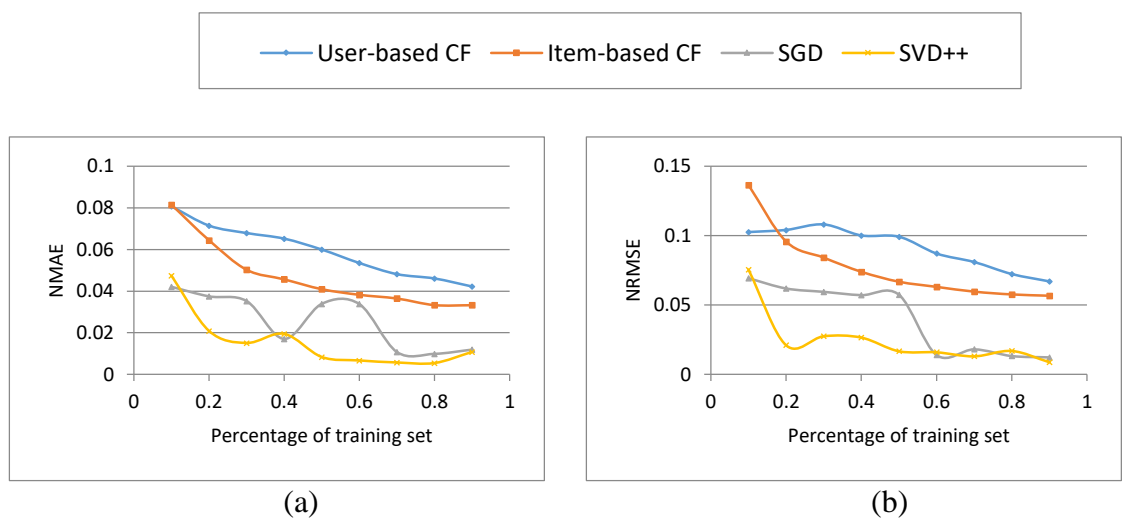

(a)                                    (b)

**Figure 6.** Performance comparison of different recommendation algorithms at different training ratios

We tested another model and it showed no improvements during evaluation. In that model, we took each user's references to a specific venue in a single year and compared those references to the total added references for that user that year. This method would show the importance of a particular venue to a user in a year, but we found that other factors, such as venue size, dominated in practice.

The use of explicit data, such as favorites or ratings for references, could improve the accuracy of recommendations. Explicit data of this nature show how interested a researcher is in an article in a much less ambiguous way. In this regard, CiteULike provides two optional but important fields that can affect venue ratings. The first field is a researcher's explicit rating of an article, and the second field is the priority a researcher has assigned to reading an article. These explicit ratings would improve PVR measurements, especially for researchers who have an interest in smaller-size venues. However, in order to collect data pertaining to these two fields, it would be necessary to construct a new dataset. Our current dataset contains unique article IDs, rated only by the first researcher who added the article to CiteULike.

## 6. Conclusion and Future Work

Multidisciplinary research areas are growing at a tremendous rate, and the number of scholarly venues is increasing every year. Researchers need to discover venues that are of interest to them, and research institutions need to be aware of these venues. In this paper, using data from an academic social network, we described an approach to recommend scholarly venues for researchers to follow and/or to publish their work in based on their current interests.

We developed a new weighting strategy for rating venues based not only on personal references, but also on the temporal factor of when the references were added. Of the similarity metrics, Euclidean-weighting achieved the best precision, recall, and NDCG. Of the recommendation algorithms, SVD++ achieved the lowest error rates. Our experiments with this strategy using a real dataset produced results that showed improvements in accuracy and ranking quality compared with a standard baseline. A number of factors will be investigated to improve the results and recommendation quality, including the total number of papers published in a venue, the number of online references to a venue in an academic social network, the average number of references added by researchers to an online reference management system, the dates on which references were added to the researchers' repositories, and the readership statistics for an article.

In future research, we plan to enhance the quality of our generated recommendations by using a researcher's trustworthiness and reputation (Alhoori, Alvarez, Furuta, Miguel Mu, & Urbina, 2009), cited references (Thor, Marx, Leydesdorff, & Bornmann, 2016), and various altmetrics (Thelwall, Haustein, Larivière, & Sugimoto, 2013) with the goal of improving accuracy, diversity, novelty, and serendipity (Ge, Delgado-Battenfeld, & Jannach, 2010). Also planned is a user study through which we will collect explicit ratings to compare with our implicit ratings. The system will begin similarly, using metadata of articles, such as title, abstract, keywords, and tags, to recommend venues, but will diverge into an analysis of explicit user-provided ratings. These experiments will use a hybrid approach implementing both collaborative filtering and content-based filtering. In addition, other factors that affect researchers' choices will be considered, such as budget availability and the ability to travel in cases such as conferences or workshops.

**Acknowledgements:** This publication was made possible by NPRP grant # 4-029-1-007 from the Qatar National Research Fund (a member of Qatar Foundation). The statements made herein are solely the responsibility of the authors. This work was supported in part by the Office of Advanced Scientific Computing Research, Office of Science, U.S. Department of Energy, under Contract DE-AC02–06CH11357.